\documentclass[3p,times,twocolumn]{elsarticle}

\usepackage{ecrc}

\volume{00}

\firstpage{1}

\journalname{Nuclear Physics B Proceedings Supplement}

\runauth{}

\jid{nuphbp}

\jnltitlelogo{Nuclear Physics B Proceedings Supplement}

\usepackage{amssymb}

\usepackage[figuresright]{rotating}

\usepackage{bbm}
\usepackage{footmisc}
\usepackage{hyperref}
\usepackage{natbib}
\usepackage{amsmath}
\usepackage{caption}
\usepackage{subcaption}
\usepackage{xspace}
\usepackage{slashed}

\newcommand{\fig}[1]{Fig.~\ref{#1}}

\newcommand{\wwjj}{\ensuremath{W^\pm W^\pm jj~}}
\newcommand{\wajj}{\ensuremath{W^\pm \gamma jj~}}
\newcommand{\wzjj}{\ensuremath{W^\pm Z jj~}}
\newcommand{\zzjj}{\ensuremath{Z Z jj~}}
\newcommand{\zajj}{\ensuremath{Z \gamma jj~}}

\newcommand{\wwjjn}{\ensuremath{W^\pm W^\pm jj}}
\newcommand{\wajjn}{\ensuremath{W^\pm \gamma jj}}
\newcommand{\wzjjn}{\ensuremath{W^\pm Z jj}}

\newcommand{\zajjn}{\ensuremath{Z \gamma jj}}

\newcommand{\order}[1]{\mathcal{O}\!\left(#1\right)}
\newcommand{\bib}[1]{Ref.~\cite{#1}}

\newcommand{\GeV}{\ensuremath{\,\mathrm{GeV}}\xspace}

\begin{document}

\begin{frontmatter}



\dochead{}

\title{%
\vspace*{-5cm}
\begin{flushright}%
{\scriptsize %
FTUV-14-1013, HU-EP-14/37, IFIC/14-64, KA-TP-28-2014, LPN14-117, MPP-2014-365,SFB/CPP-14-76}
\end{flushright}
\vspace*{4.4cm}
QCD Induced Di-boson Production in Association with Two Jets at NLO QCD}


\author[a]{Francisco Campanario\fnref{1}}
\author[b,c]{Matthias Kerner}
\author[b,d]{Le Duc Ninh}
\author[b]{Dieter Zeppenfeld}
\fntext[1]{Speaker. Prepared for the 37th International Conference on High Energy Physics (ICHEP 2014), 
2-9 Jul 2014, Valencia, Spain.}
\address[a]{Theory Division, IFIC, University of Valencia-CSIC, E-46980
  Paterna, Valencia, Spain}
\address[b]{Institute for Theoretical Physics, KIT, 76128 Karlsruhe, Germany}
\address[c]{Max Planck Institute for Physics, F\"ohringer Ring 6, D-80805
M\"unchen, Germany}
\address[d]{Humboldt-Universit\"at zu Berlin, Institut f\"ur Physik, 
Newtonstra{\ss}e 15, D-12489 Berlin, Germany}

\begin{abstract}
We discuss results for di-boson plus two jets
production processes at the LHC at NLO QCD. Issues related
to the scale choice are reviewed. We focus on the distributions of the invariant mass 
and rapidity separation of the two hardest jets and show, for \wajj and \zajj production, 
how the contribution from the radiative decays of the massive gauge bosons can
be significantly reduced.
\end{abstract}


\end{frontmatter}


\section{Introduction}
\label{sec:intro}
The experimental program at the LHC for measuring the di-boson in association
with two jets production processes has started. Results for the same-sign \wwjj
production process have already been reported by the ATLAS and CMS
collaborations~\cite{Aad:2014zda,CMS:2014uib}. They have presented first
evidence for the electroweak (EW) induced production mechanism, thus, being able to
distinguish it from the QCD induced one, considered to be a background, in the
framework of vector boson scattering and quartic gauge coupling measurements.

Generally, the electroweak-induced production mechanism of order
$\order{\alpha^ 4}$ (for on-shell vector boson production) can be classified
into the $t$-channel vector-boson contributions, known for all the processes
at NLO
QCD~\cite{Jager:2006zc,Jager:2006cp,Jager:2009xx,Bozzi:2007ur,Denner:2012dz,Campanario:2013eta}
and other contributions, mainly, tri-boson production processes, with a
subsequent hadronic decay from one of the vector bosons. The NLO QCD
corrections are available via the {\texttt{VBFNLO}}
program~\cite{Arnold:2008rz,Arnold:2011wj,Baglio:2014uba}. They make use of
the matrix elements computed first for the tri-boson production processes
including leptonic
decays~\cite{Hankele:2007sb,Campanario:2008yg,Bozzi:2009ig,Bozzi:2010sj,Bozzi:2011wwa,Bozzi:2011en},
and involve some approximations~\cite{Feigl:2013naa}.

The NLO QCD corrections for the QCD-induced production mechanism of order
$\order{\alpha_s^2\alpha^2}$ (for on-shell production) for all the di-boson
production processes have been recently
completed~\cite{Melia:2010bm,Melia:2011dw,Greiner:2012im,Campanario:2013qba,
Campanario:2013gea,Gehrmann:2013bga,Badger:2013ava,Campanario:2014dpa,Bern:2014vza,
Campanario:2014ioa,Campanario:2014wga}. To
this programme, including the leptonic decays of the vector bosons and all
off-shell and spin-correlation effects, we have contributed with predictions
for the \wzjjn, \wajjn, \wwjjn, \zzjj and \zajj processes and the codes are available in
the {\texttt{VBFNLO}} program package. We refer to them from now on by the on-shell production
names for simplicity. In these proceedings, we briefly discuss
them. A sketch of the calculations is given in
Sect.~\ref{sect:setup}. Numerical results are presented in
Sect.~\ref{sect:num}. Finally, we conclude in Sect.~\ref{sect:con}

\section{Calculational Setup}
\label{sect:setup}

To compute the di-boson production processes in association with two jets at
NLO QCD, we follow the spinor-helicity amplitude
method~\cite{Hagiwara:1988pp,Campanario:2011cs} and the effective current
approach, factorizing, in this way, the leptonic tensor containing the EW
information from the QCD part. Two generic amplitudes contribute,
\begin{align}
  \label{eq:amp} &pp\rightarrow V_1 V_2 jj + X, \\ &pp\rightarrow \hat{V} jj +
X\;\; (\text{absent in }\wwjj).
\end{align} In each process, the leptonic decays  of the vector bosons are
included via effective currents, e.g., for \zzjj production, we have $V_i =
Z/\gamma^* \to l_i^+ l_i^-$ ($i = 1,2$) and $\hat{V}=Z/\gamma^*\to l_1^+ l_1^- l_2^+ l_2^-$.
In this way, we take into account all off-shell effects and
spin correlations. Since the leptonic tensors are globally set in our code, this
procedure makes it straightforward to implement and check all the processes. For
each process, we cross-check the LO and real emission corrections against
Sherpa~\cite{Gleisberg:2008ta,Gleisberg:2008fv} and agreement 
is found for integrated cross sections for all processes. Additionally,
we have implemented two completely independent calculations for the
processes~(see \bib{Campanario:2013qba}).

For processes involving photons in the final state, namely \wajj and \zajjn,
several technicalities arise: (1) a new set of scalar integrals not present in
the off-shell photon case appears. We have checked the scalar-integral basis using two
independent calculations; (2) to
avoid the need of including photon fragmentation functions and to preserve the
exact cancellation of the QCD infrared singularities, we use the photon
isolation criterion \`a la Frixione~\cite{Frixione:1998jh}; (3) to optimize the
Monte Carlo integration efficiency, the phase-space generator is divided into
two separate regions generated as double EW boson production, $V_1 \gamma$
with subsequent decay of the $V_1$ vector boson, as well as $\hat{V}$
production with three-body decays
$\hat{W}(\hat{Z})\to l\nu (l^+l^-,\bar{\nu}\nu) \gamma$ (see
\bib{Campanario:2014ioa} for details).

At the partonic level, we classify the amplitudes into sub-processes with 4
quarks and those with 2 quarks and 2 gluons. The latter is absent in the
\wwjj production process. This fact makes this channel very interesting since
interference effects between the QCD- and the EW-production mechanisms are
expected to be maximal. 

Six-point rank-five one-loop tensor integrals appear in the 
$2$-quark-$2$-gluon virtual amplitudes. For the \zzjj
production process, there are up to $42$ six-point diagrams in the $gg\to u
\bar u ZZ$ sub-amplitude. The $4$-quark group is simpler with up to $24$
generic hexagons for the subprocess $q_1q_2\to q_1q_2 ZZ$. Since the
kinematics are the same when replacing a Z boson for a photon, the hexagons can be
re-used, using a cache system, in the other contributing sub-amplitudes, i.e.,
$Z\gamma^*jj$, $\gamma^* Zjj$ and $\gamma^* \gamma^*jj$. 

Additionally, there are closed quark-loop diagrams. For the neutral
production processes, we do not include closed-quark loops where the vector
bosons or/and the Higgs boson are directly attached to the loop. This set of
diagrams forms a gauge invariant subset and contributes at the few per mille
level to the NLO results~\cite{Greiner:2012im}, and hence is negligible for
all phenomenological purposes. The diagrams with a closed quark-loop
and two or three gluons attached to it are however included.

With our program, we obtain the NLO inclusive cross section with statistical
error of $1\%$ within 20 minutes (\wwjjn) to 4 hours (\zajjn) on an
 Intel $i7$-$3970X$ computer with one core
and using the compiler Intel-ifort version $12.1.0$. The distributions shown
below are based on multiprocessor runs with a total statistical error of
0.03\% at NLO. %
\begin{figure}[ht!]
  \centering
  \includegraphics[width=0.9\columnwidth]{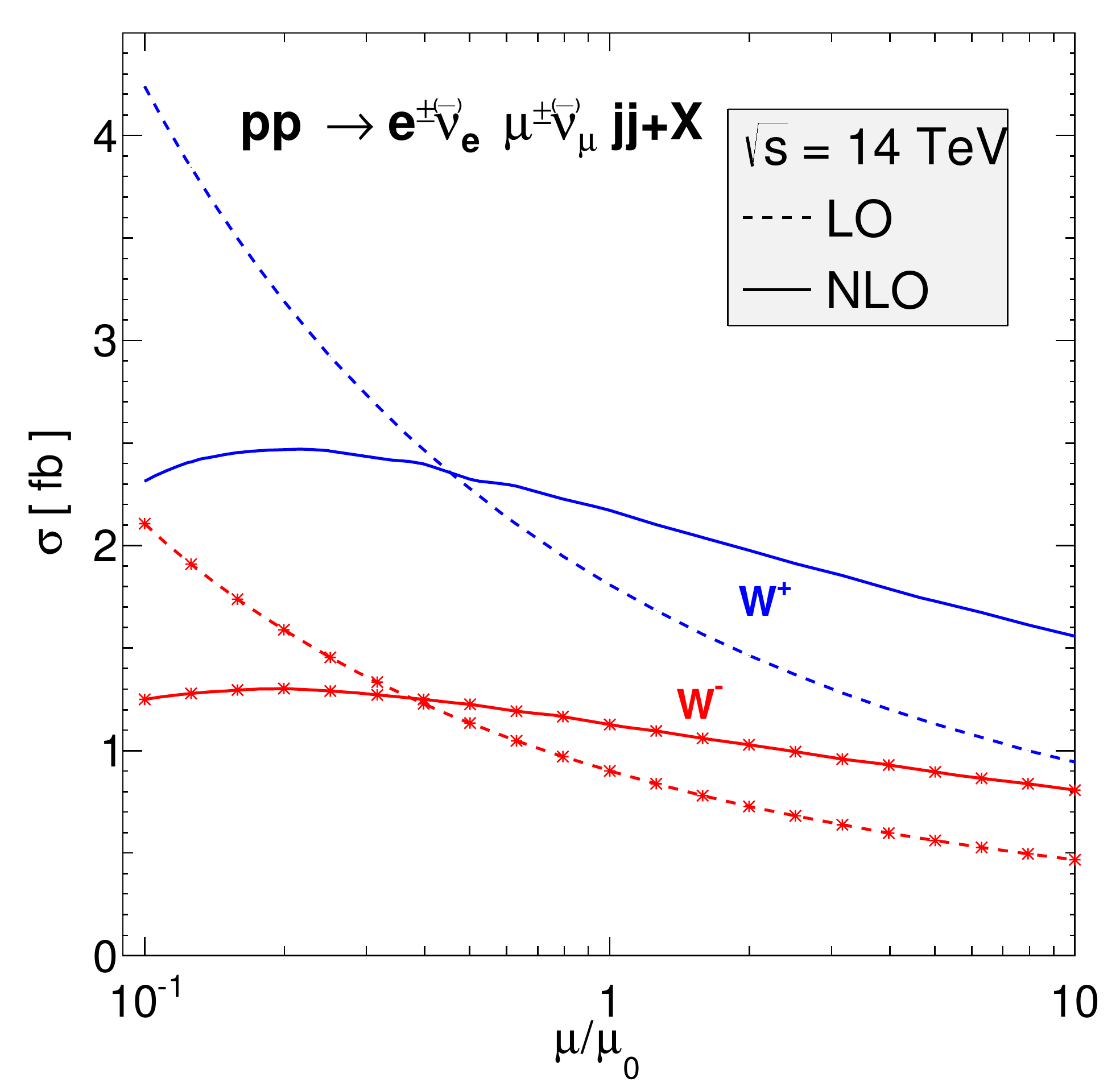}
  \caption{Scale dependence of the LO and NLO cross sections at the LHC. 
The curves with and without stars are for $W^-W^-jj$ and $W^+W^+jj$
productions, respectively. 
The cuts used are defined in the text.}
  \label{fig:scale}
\end{figure}
\begin{figure*}[ht!]
  \centering
  \includegraphics[width=0.95\columnwidth]{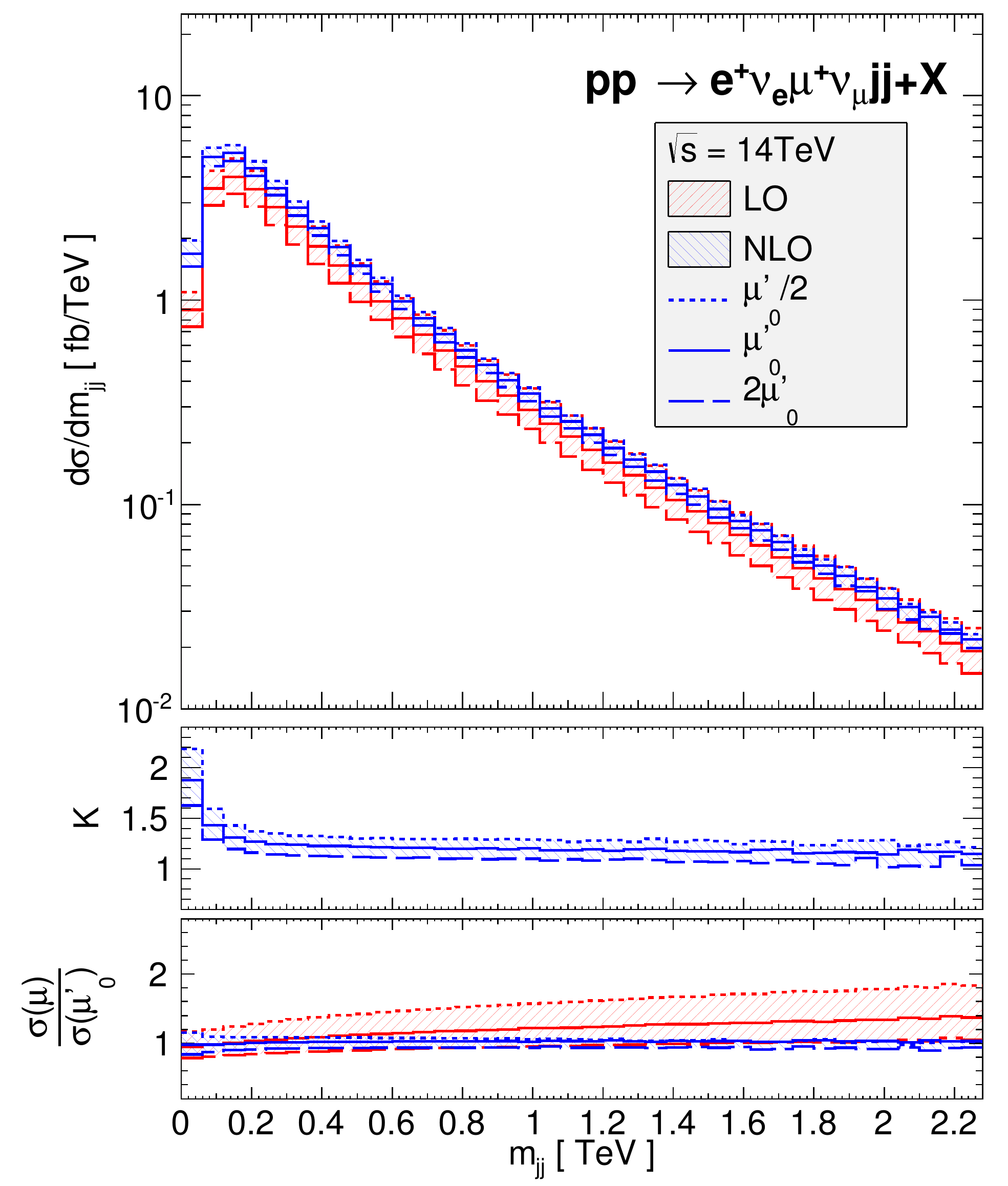} \hfill
  \includegraphics[width=0.95\columnwidth]{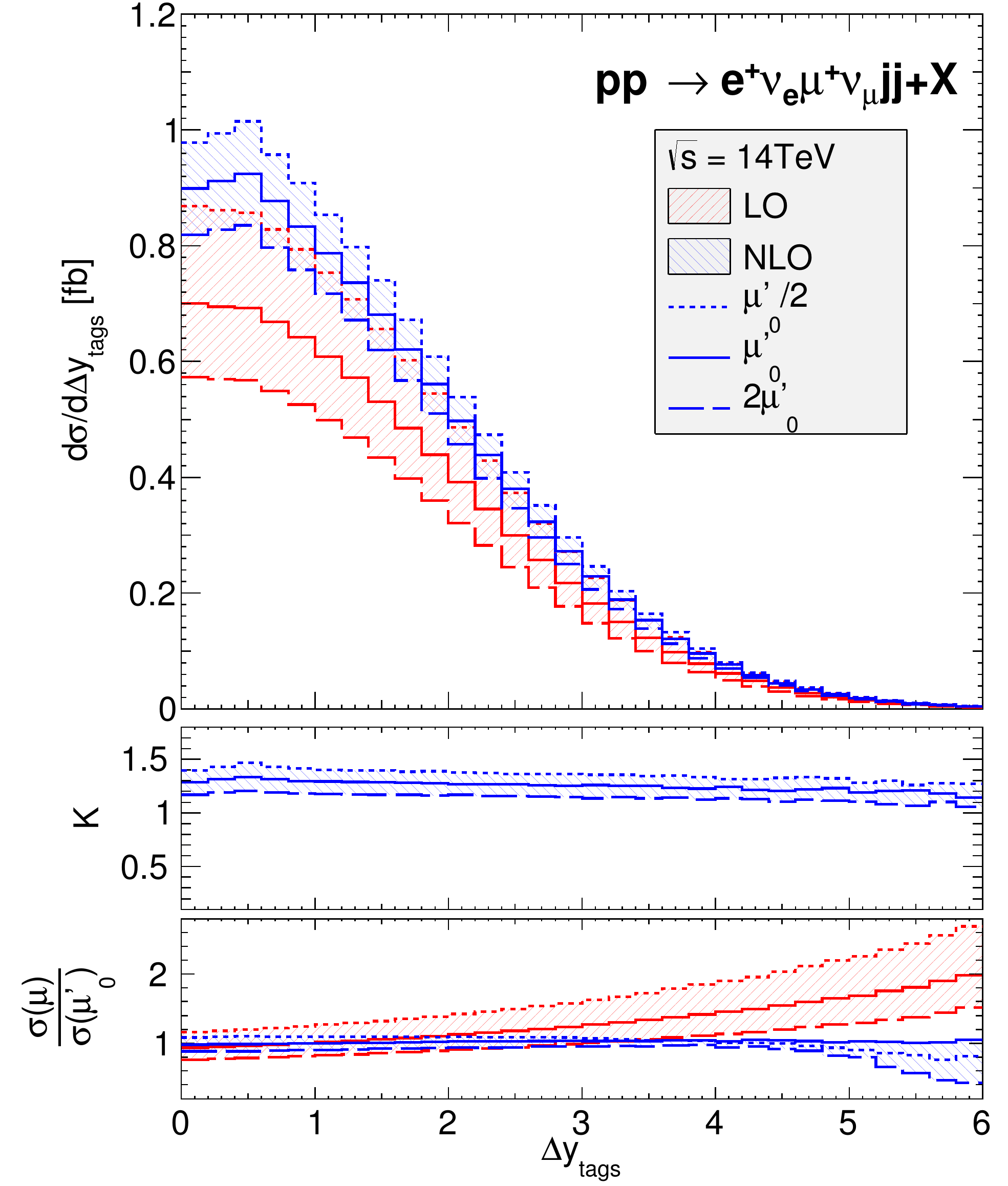}
  \caption{Differential cross sections for the QCD-induced channels at LO and
    NLO for the rapidity separation between the two tagging jets (right) and its
    invariant mass (left). The bands for the upper and middle panel describe 
$\mu_0^{\prime}/2 \le \mu_F=\mu_R\le 2\mu_0^{\prime}$ variations. The
$K$-factor bands 
are due to the scale variations of the NLO results, 
with respect to $\sigma_\text{LO}(\mu_0^{\prime})$. 
The solid lines are for the central scale while the dotted and dashed lines
correspond
 to $\mu_0^{\prime}/2$ and $2\mu_0^{\prime}$, respectively.
Additional panels at the bottom show the ratios of differential 
cross sections with the scale $\mu_{0}$ over the ones with $\mu^{\prime}_{0}$ at 
LO and NLO. The bands on these ratios 
show the scale variations $\mu_0/2 < \mu < 2 \mu_0$ of the numerators while
 the denominators are calculated at 
the central scale. 
}
\label{dist_NLO_jets_inc_prime}
\end{figure*}
\section{Numerical Results}
\label{sect:num}
In the following, we present results for the LHC run at 14 TeV. We use the
anti-kt algorithm~\cite{Cacciari:2008gp} and consider jets that lie in the
rapidity range $|y_{\text{jet}}|< 4.5$ and have transverse momenta
$p_{T,\text{jet}}> 20 $ GeV with a cone radius of R=0.4. For leptons, we use
\begin{equation} p_{T,l} > 20 \GeV \quad |y_l| < 2.5 \quad R_{lj(l)} > 0.4,
\end{equation} and for processes with an on-shell photon,
\begin{equation} p_{T,\gamma} > 30 \GeV ~~~~|y_\gamma| < 2.5\quad R_{\gamma
j(l)} > 0.7(0.4),
\end{equation} with the smooth isolation criterion \`a la Frixione~\cite{Frixione:1998jh}. With a cone
radius of $\delta_0 = 0.7$, events are accepted if
\begin{equation} \sum_{i\in \text{partons}} p_{T,i}\theta(R-R_{\gamma i}) \le
p_{T,\gamma}\frac{1-\cos R}{1-\cos\delta_0} \quad \forall R<\delta_0.
\label{eq:Frixione cut}
\end{equation} Finally, for processes with the $W$ bosons, we impose that the missing
transverse momentum associated with the neutrinos is $\slashed{p}_T > 30 \GeV$. 
For \wzjj we have $m_{l^+l^-} > 15 \GeV$ in addition.

As EW input parameters, we use $M_W=80.385\GeV$, $M_Z=91.1876\GeV$ and $G_F=1.16637
\times 10^{-5}\GeV^{-2}$ and derive the mixing angle and the electromagnetic
coupling from tree level relations. We use the MSTW2008 parton distribution
functions~\cite{Martin:2009iq} with $\alpha^{LO(NLO)}_S (M_Z)=0.13939
(0.12018)$. We assume a unit CKM matrix and consider all fermions massless, except
the top quark with $m_t=173.1 \GeV$. The decay widths are fixed at $\Gamma_W = 2.09761 \GeV $ and at
$\Gamma_Z = 2.508905 \GeV $. We use the five flavor scheme. The top-quark contribution is decoupled from the
running, but is explicitly included in the one-loop amplitudes.

We only consider equal renormalization and factorization scales in the following, but allow for three different choices of the central scales:
\begin{align}
  &\mu_{0}=\frac{1}{2}
\left(\sum_{\text{partons}} p_{T,i} + 
\sum_{V_i}\sqrt{p_{T,V_i}^2+m_{V_i}^2}\right), \nonumber
\end{align}
\begin{align}
 &\mu^{\prime}_{0}=\frac{1}{2}
\left(\sum_{\text{jets}} p_{T,i} \exp{|y_{i}-y_{12}|} + 
\sum_{V_i}\sqrt{p_{T,V_i}^2+m_{V_i}^2}\right),\nonumber\\
 & \mu_{0}^{\prime \prime}=\frac{1}{2} \left[E_{T}(jj) +
  E_{T}(VV)\right],
\label{eq:scale}
\end{align}
with $V_i \in (W,Z,\gamma)$. $m_{V_i}$ denotes the invariant
mass of the corresponding leptons ($m_{V_i}=0$ for on-shell photons) and $y_{12} = (y_1 +
y_2)/2$ the average rapidity of the two hardest (or tagging) 
jets, ordered by decreasing transverse momenta.  $E_{T}(jj)$ and
$E_{T}(VV)$ stand for the transverse energy of the two tagging jets and of the
$VV$ system, respectively. In the last two scale choices of Eq.~\eqref{eq:scale}, the first term
interpolates between $m_{jj}$ and $\sum p_{T,jets}$ for large and small
$\Delta y_{jj}=|y_1-y_2|$ values, characterizing the dynamics of these processes
appropriately, as we will see below.

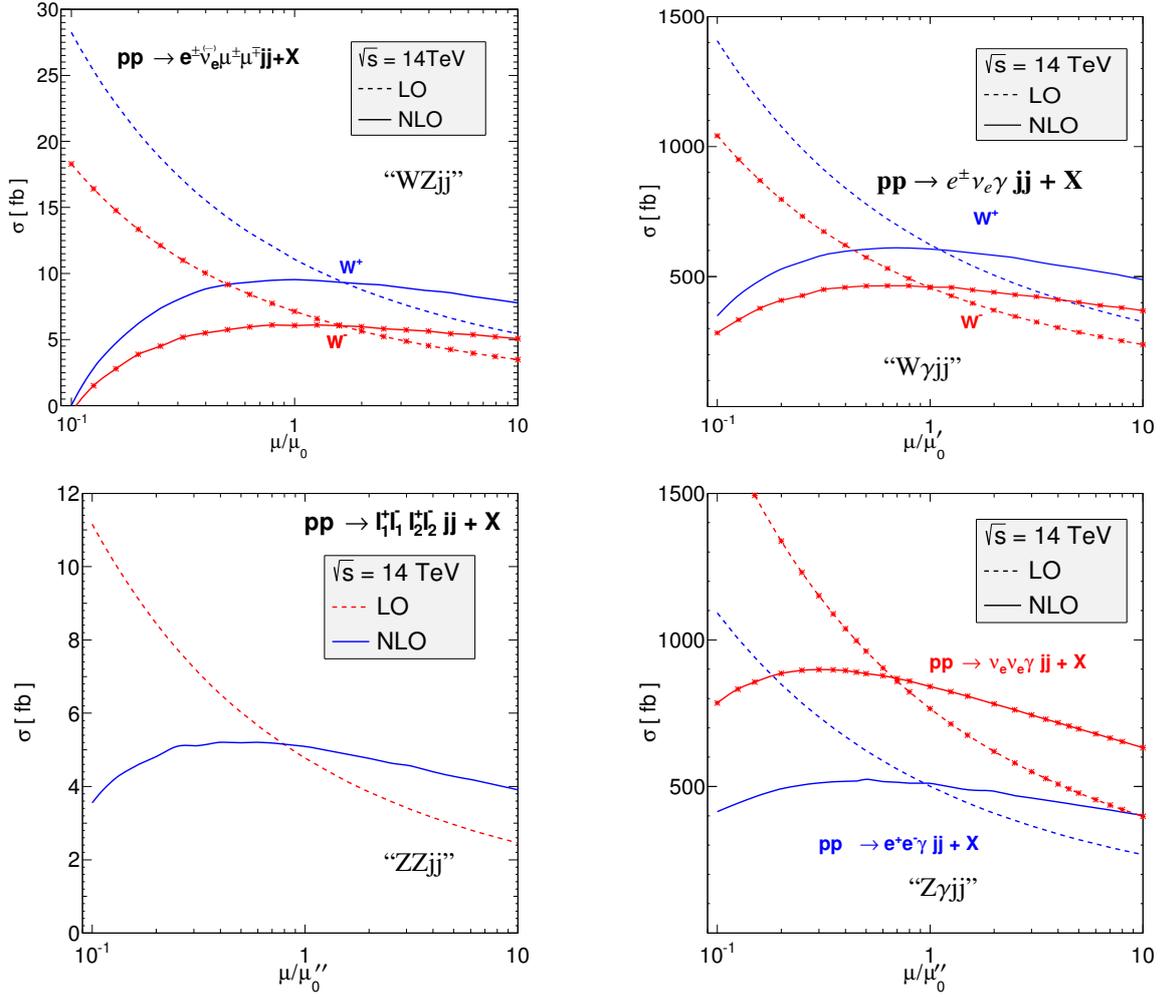
\begin{figure*}[ht!]
  \centering
  \setlength{\unitlength}{0.1\textwidth}
  \begin{picture}(8,8)(0.8,0)
\put(0,4.3){\includegraphics[width=0.43\textwidth]{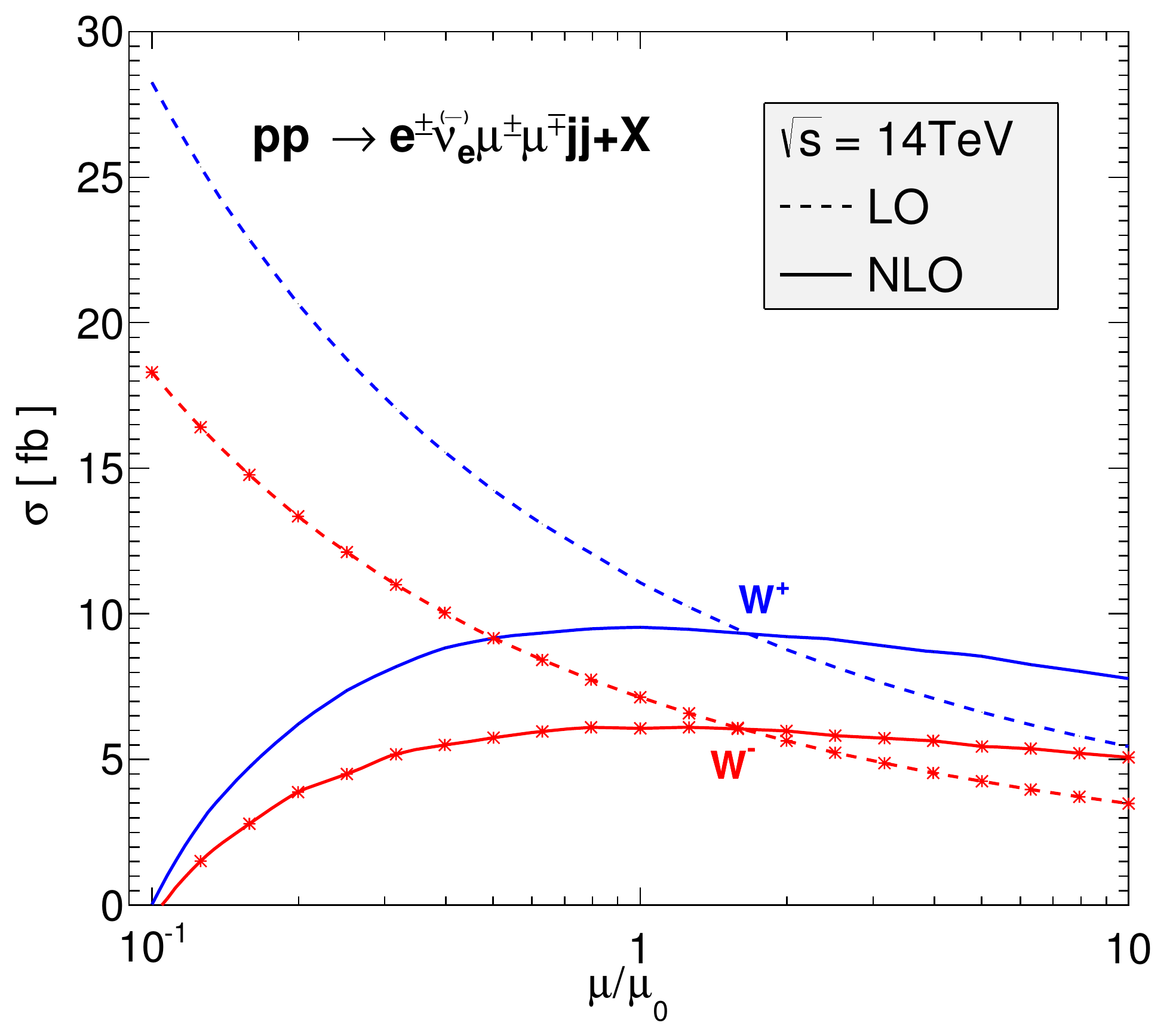}}
\put(5,4.3){\includegraphics[width=0.43\textwidth,height=0.37\textwidth]{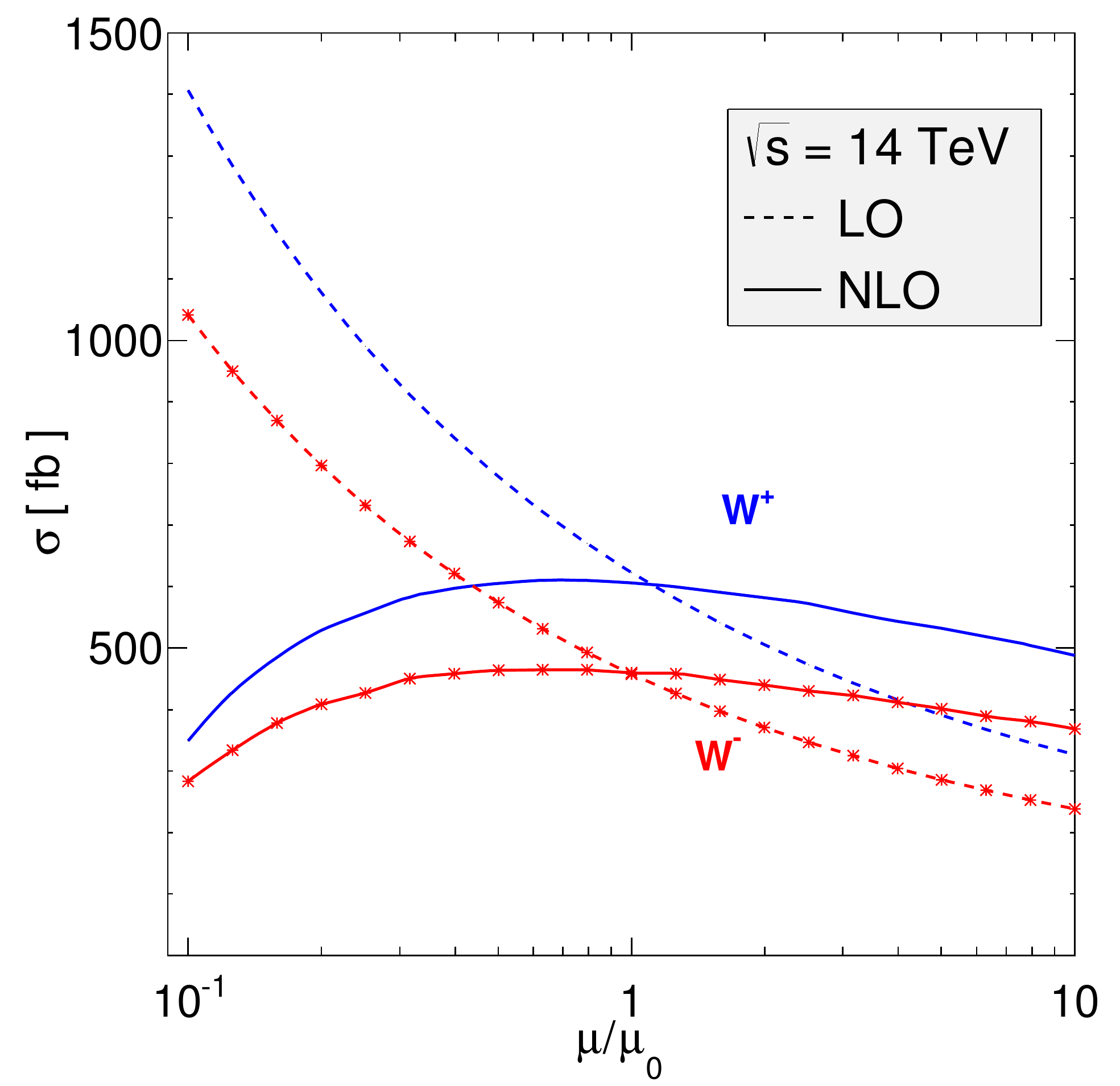}}
\put(0,0){\includegraphics[width=0.43\textwidth]{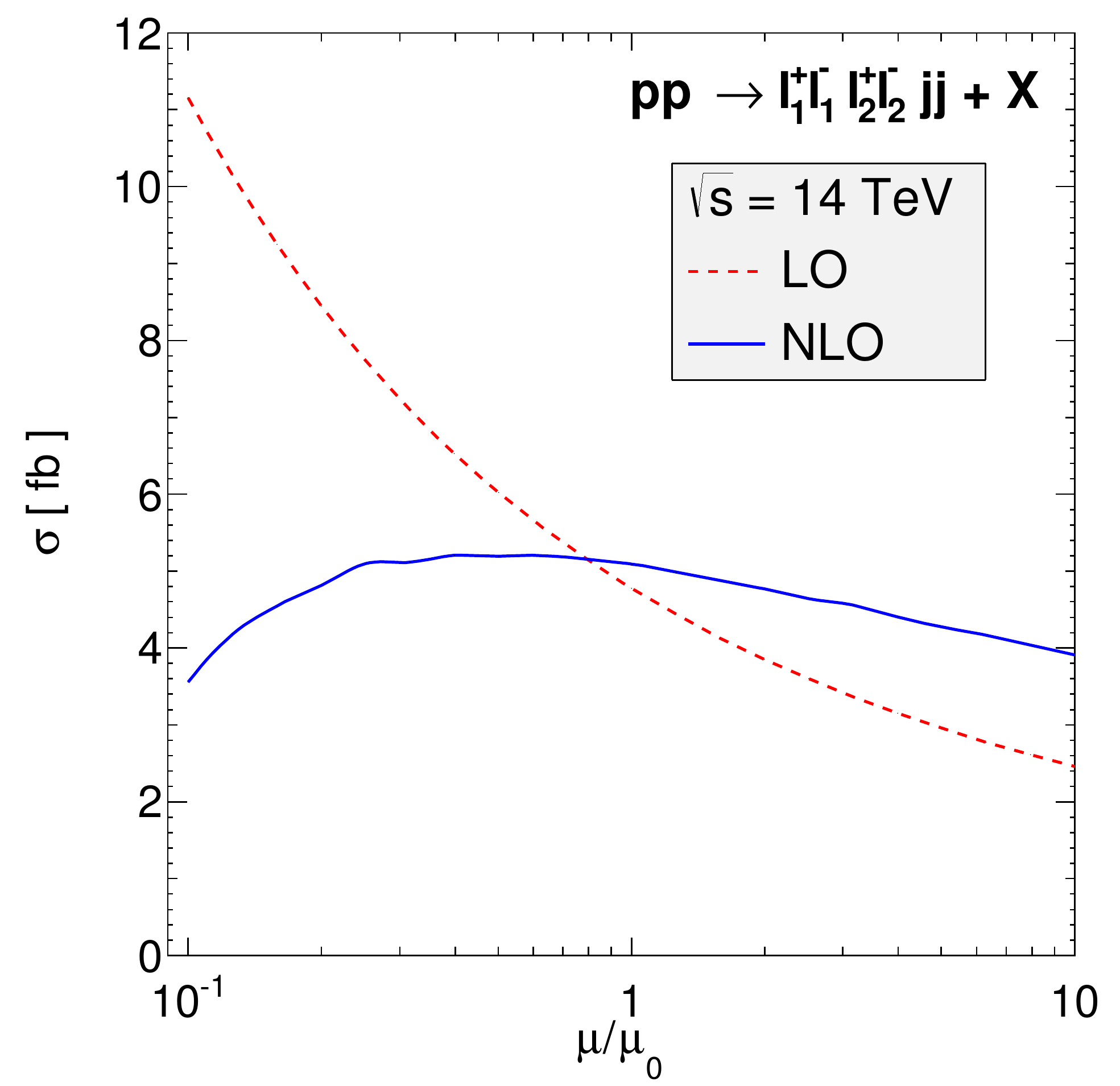}}
\put(5,0){\includegraphics[width=0.43\textwidth]{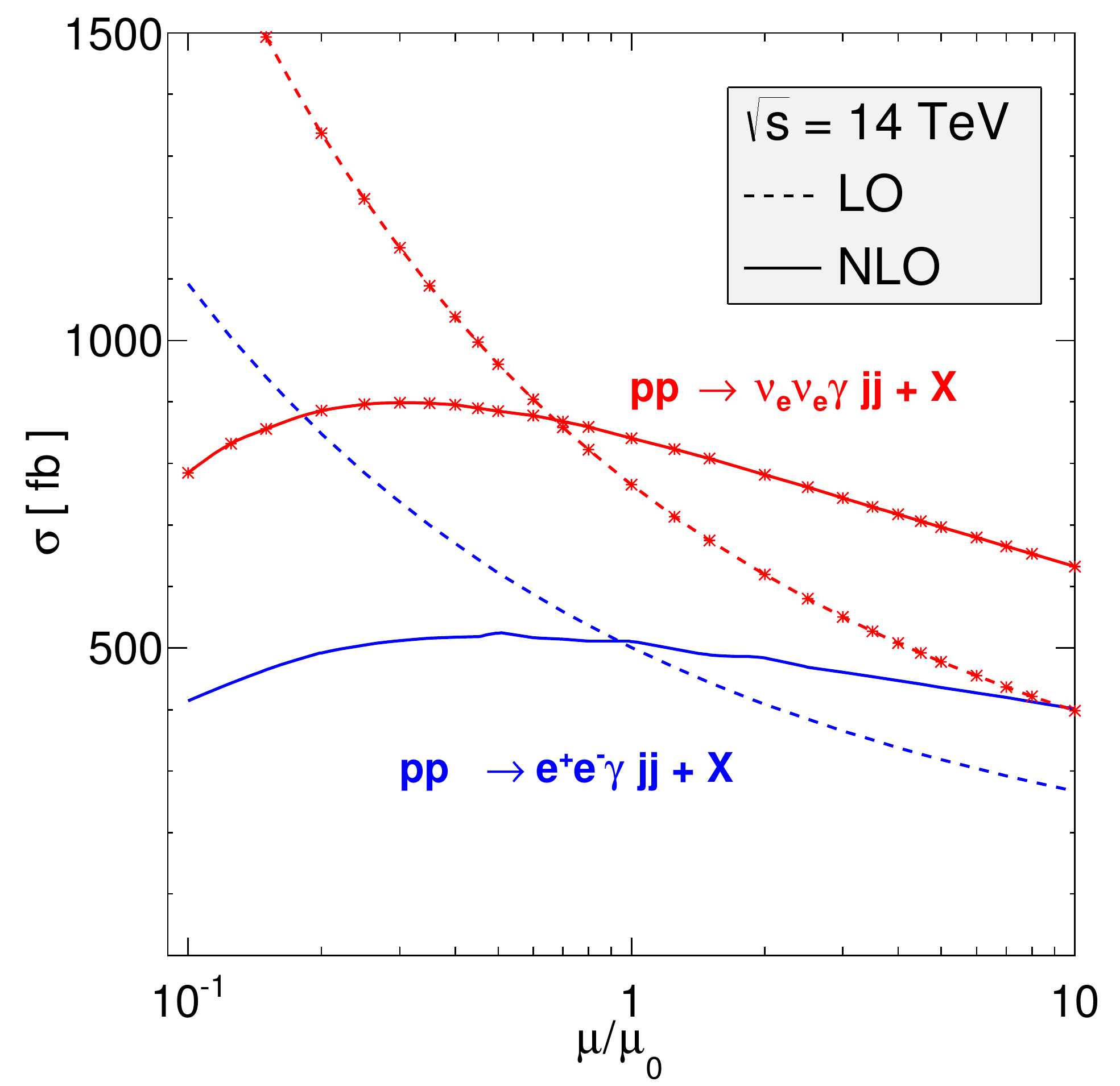}}
\put(7,6.5){{\bf  pp $\rightarrow e^\pm \nu_e \gamma$ jj + X}}

\put(7,5.){{  ``W$\gamma$jj''}}
\put(7.42,4.48){{ {\scriptsize$\prime$}}}
\put(7.2,0.8){{  ``Z$\gamma$jj''}}
\put(3,6.5){{  ``WZjj''}}
\put(3,1.0){{  ``ZZjj''}}

\put(7.4,0.18){{ {\scriptsize $\prime$$\prime$}}}
\put(2.45,0.18){{ {\scriptsize $\prime$$\prime$}}}
\end{picture}
  \caption{Scale dependence of the LO and NLO cross sections at the LHC. The
    central scales are defined in Eq.~\eqref{eq:scale}.
}
\label{fig:scaleVV}
\end{figure*}
\begin{figure*}[ht!]
  \centering
  \includegraphics[width=0.95\columnwidth]{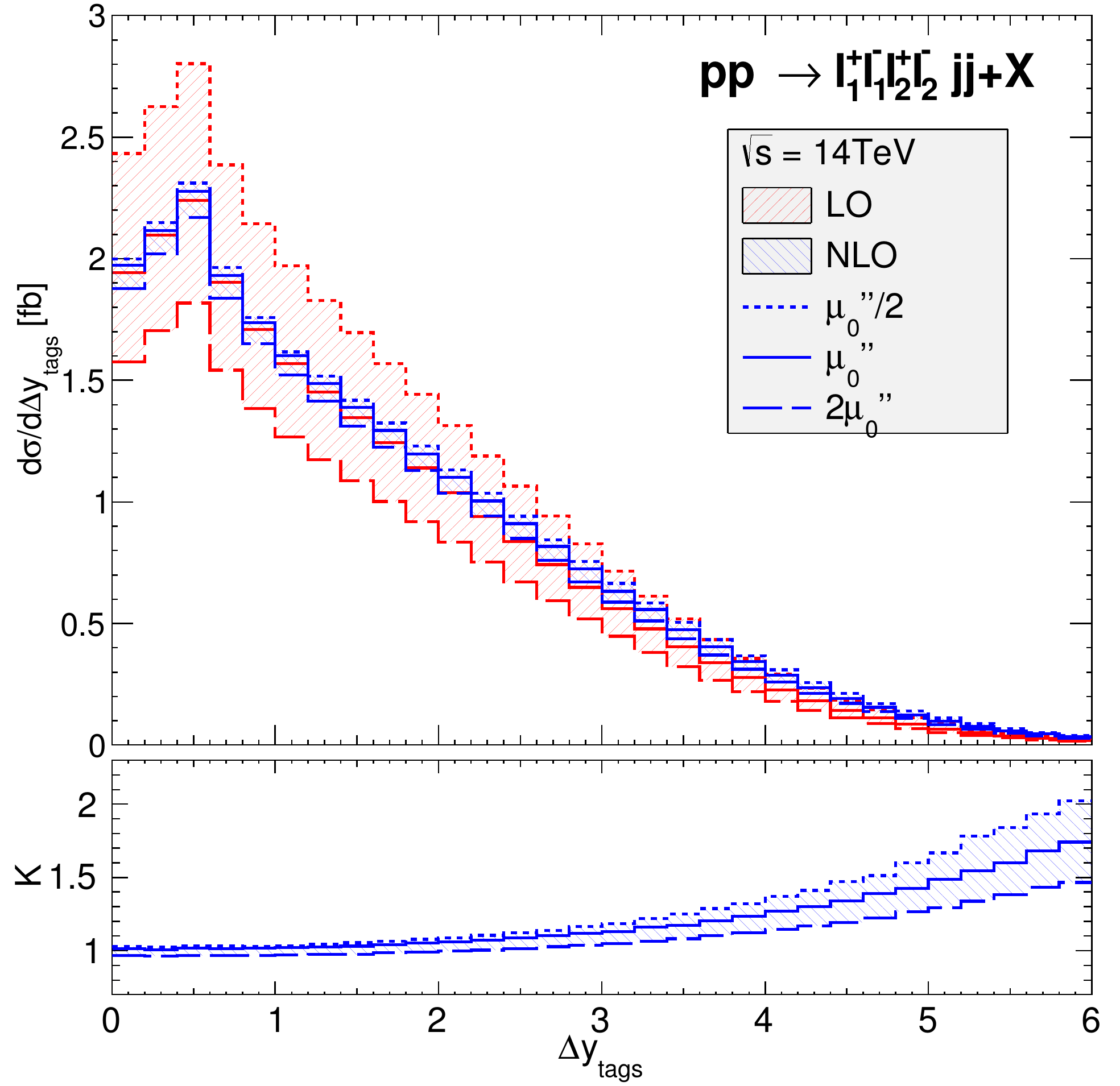}
  \includegraphics[width=0.95\columnwidth]{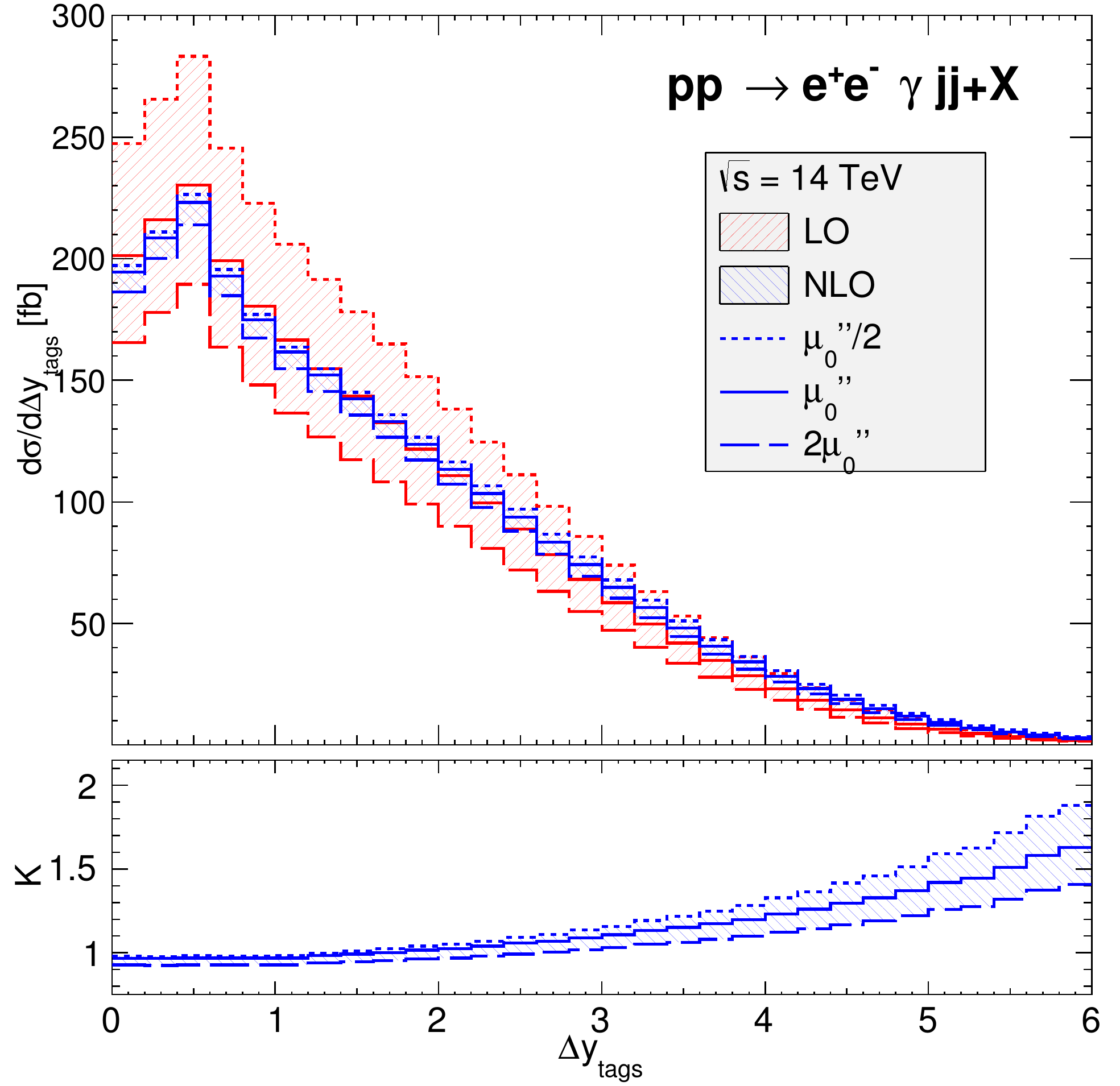}\\
  \caption{Differential cross sections for the QCD-induced channels at LO and
    NLO for the rapidity separation between the two tagging jets for the \zzjj
    (left) and the \zajj (right) production processes. More details in Fig.~\ref{dist_NLO_jets_inc_prime}.}
\label{fig:deltayZZ}
\vspace*{0.2cm}
\includegraphics[width=0.95\columnwidth]{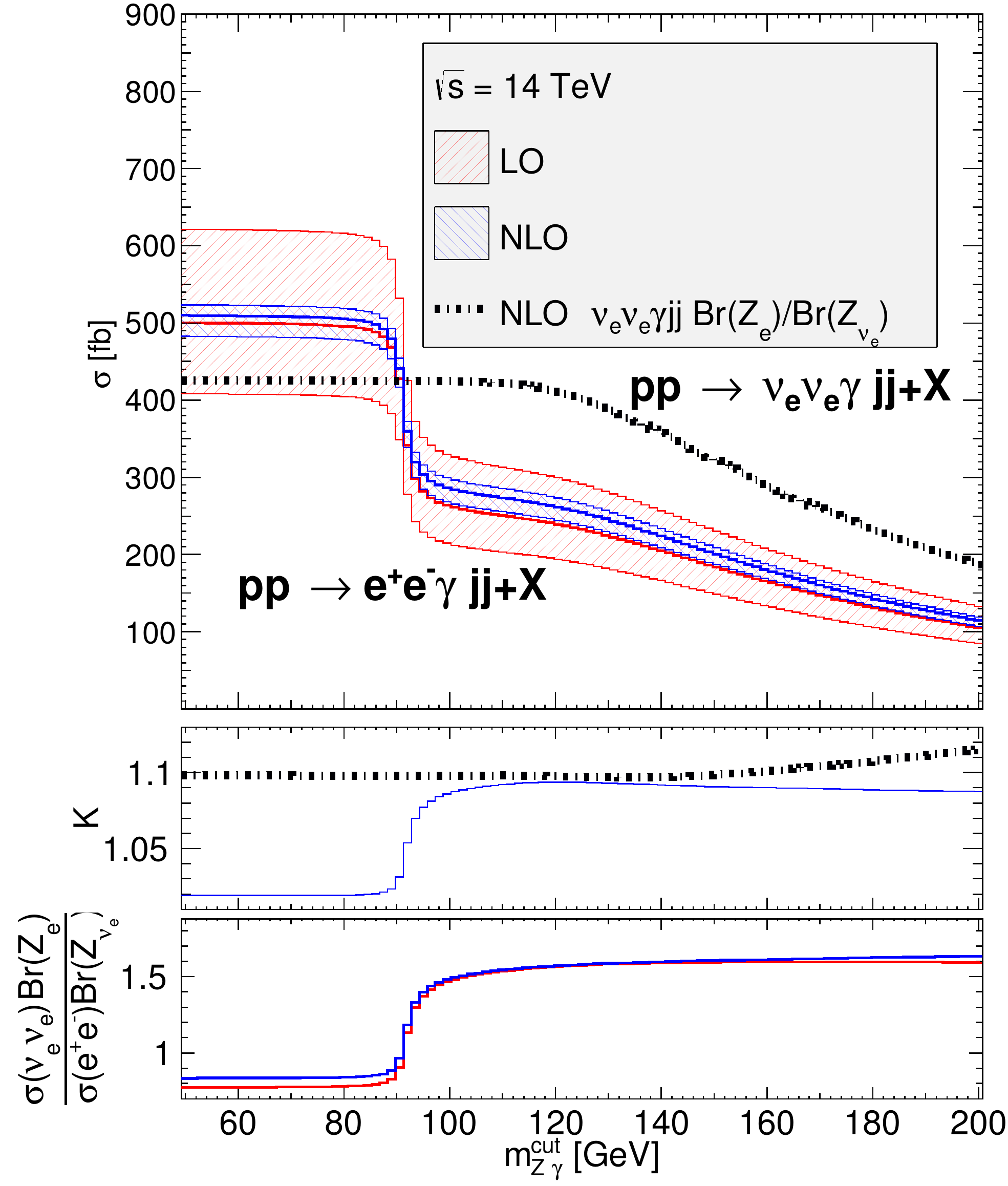}%
\hspace*{0.3cm}
  \includegraphics[width=0.95\columnwidth]{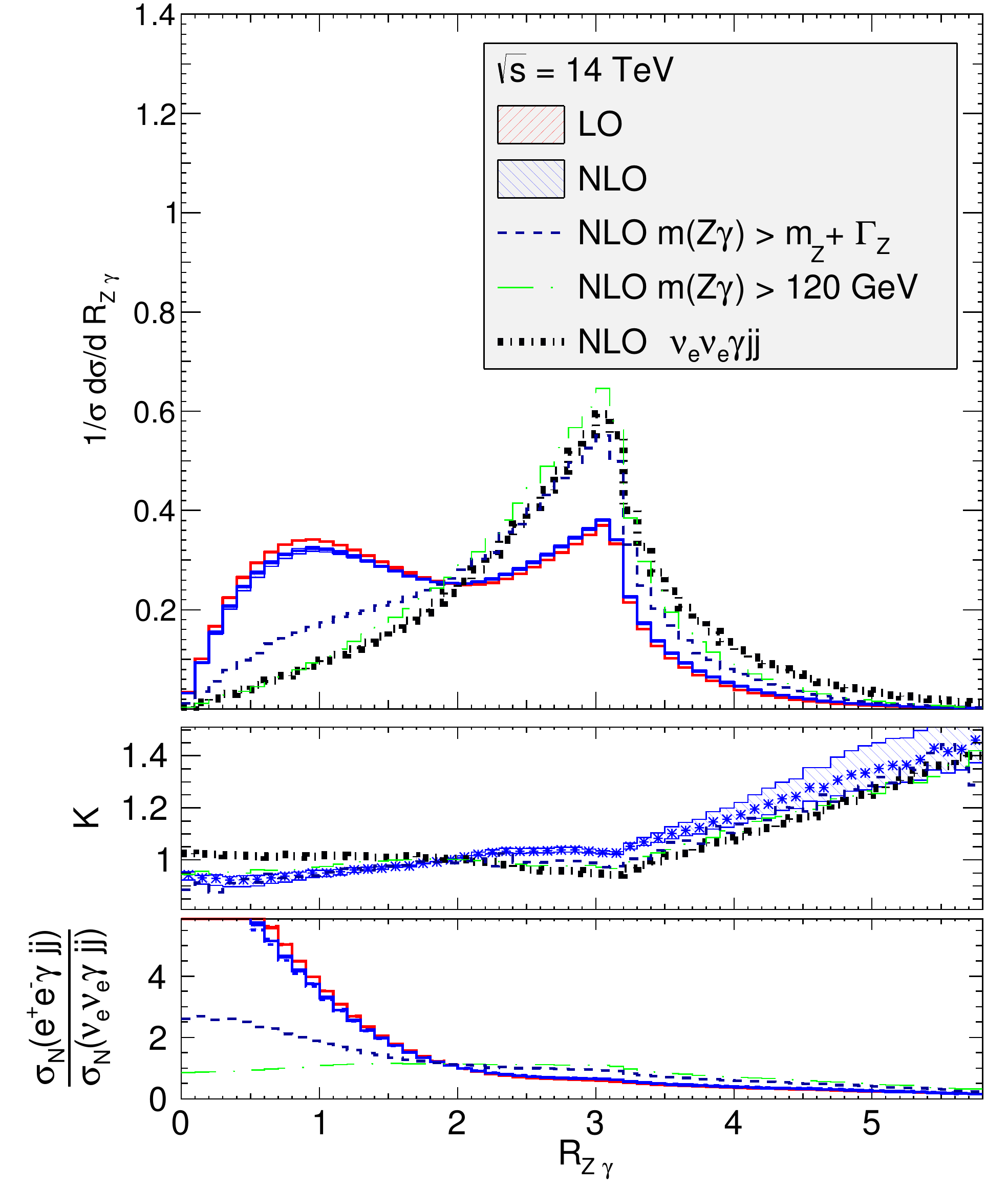}
  \caption{%
Left: Cross section for different values of the reconstructed
     $Z\gamma$ invariant mass cut. $\mu_0^{\prime\prime}$ is chosen as a central scale. The neutrino curve is multiplied by the
     ratio of the charge-lepton versus neutrino branching ratios. The middle
     panel shows the K-factor and the lower the ratios of the modified neutrino
     cross section versus the LO and NLO electron cross sections. 
     Right: Normalized differential distributions of the rapidity-azimuthal angle separation 
     $R_{Z\gamma}$ for different values of the
     $m_{Z\gamma}^{\ensuremath{\,\mathrm{cut}}}$ cut. The middle and lower panels show the
     differential K-factor plots and the ratios of the normalized electron versus
     neutrino pair production channels.
}
\label{fig:finalrad}
\end{figure*}
\begin{figure}[ht!]
  \centering
  \includegraphics[width=0.95\columnwidth]{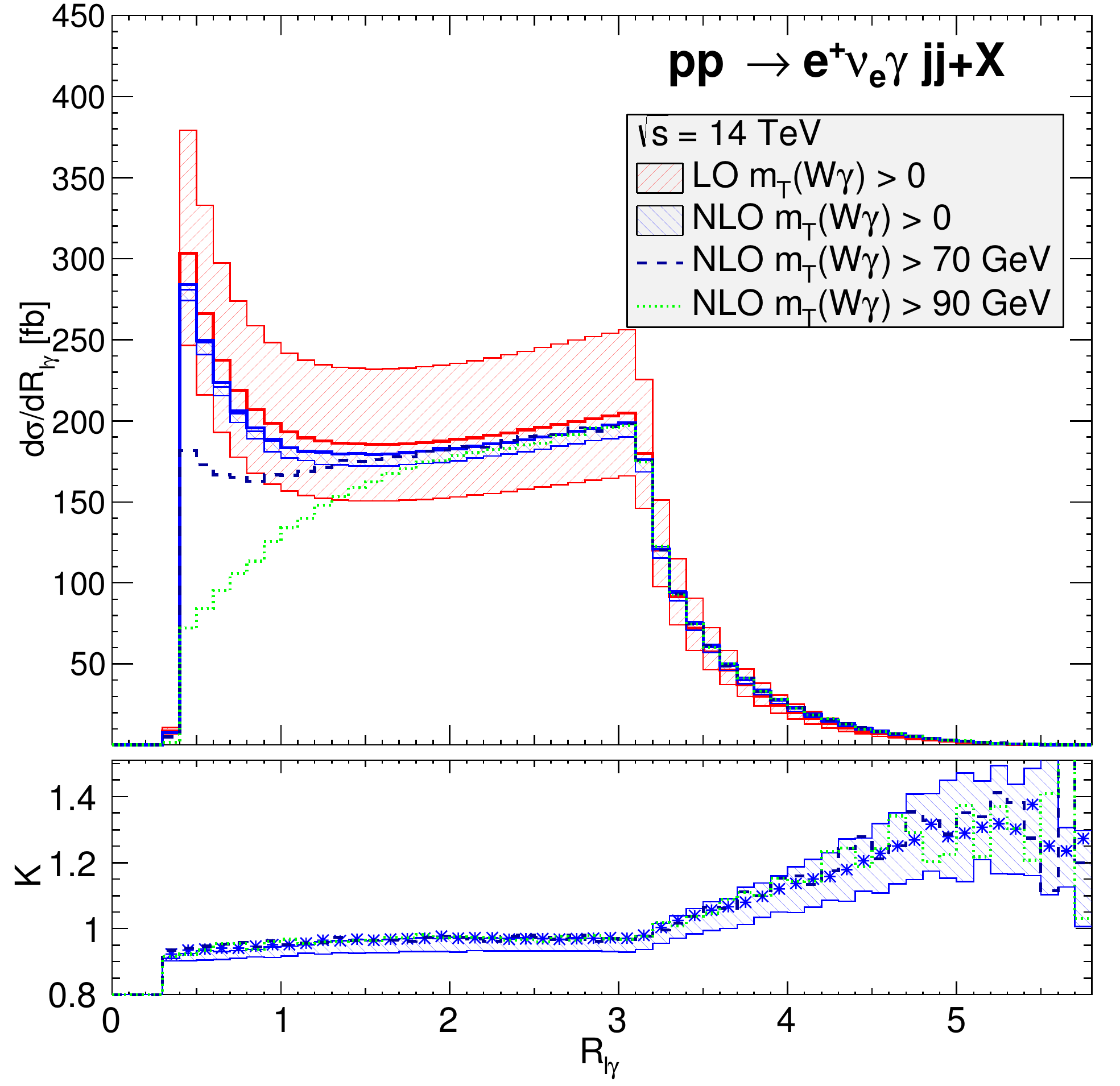}
  \caption{%
Differential cross sections of rapidity-azimuthal angle separation for different
values of the transverse cluster
energy of the $W\gamma$ system $m_{T,(W\gamma)}$. $\mu_0^{\prime}$ as a central scale is chosen.%
}
\label{fig:finalradWA}
\end{figure}
In Fig.~\ref{fig:scale}, the renormalization and factorization scale variation
plot is shown for $W^-W^-jj$ and $W^+W^+jj$ production. We observe a significant
reduction in the scale dependence around $\mu_0$ at NLO QCD. The uncertainties
obtained by varying $\mu_{F,R}$ by a factor of $2$ above and below the central value are
$45\%$ ($45\%$) at LO and $16\%$ ($18\%$) at NLO for the $W^+W^+$ ($W^-W^-$)
channel. If the two scales are varied separately (not shown), a small
dependence on $\mu_F$ is observed, while the $\mu_R$ dependence is similar to
the behavior shown in \fig{fig:scale}.

To illustrate the phase space dependence, we plot in
Fig.~\ref{dist_NLO_jets_inc_prime}, the differential distributions of the
invariant mass (left) and the rapidity separation (right) for two different
choices of the central scale. The upper and middle panel show the curves with
respect to $\mu_o^{\prime}$. The middle panel shows the K-factor, defined as
the ratio of the NLO to the LO predictions. Note the almost flat K-factor in
the whole spectrum. This is not the case if the $\mu_0$ scale is chosen as a
central scale as can be inferred from the lower panels where the ratio of the
cross sections for the two scales are plotted. Note the large differences of
order of 2 at the LO for the $\Delta y_{jj} $ distribution. This shows the
sensitivity of the LO predictions to different scale choices and the relevance
of the NLO predictions to stabilize the results. The failure of the $\mu_0$
scale to describe the dynamics can be understood in the following way: The
invariant mass $m_{jj}$ of the two leading jets rapidly increases at 
large rapidity separation $\Delta y_{jj}$, even though
the tagging jets are mainly produced with low $p_T$, as can be seen from $ m^2_{jj} \approx
2p_{T,j1}p_{T,j2}[\cosh(y_{j1}-y_{j2}) - \cos(\phi_{j1}-\phi_{j2})]$. Note the
exponential growth of $m_{jj}$ with $\Delta y$. The low value of $p_{T,j2}$
acts as a veto for further (central) jet activity, resulting in large QCD
uncertainties.
%

In Fig.~\ref{fig:scaleVV} we show the scale uncertainties
for \wzjj and \wajj in the upper row and for \zzjj and \zajj in the lower 
row. The
reduction of the scale uncertainties is similar and significant in all the
production processes going from 40$\%$ at LO to below 10$\%$ at NLO. In
Fig.~\ref{fig:deltayZZ}, the differential distribution for the
rapidity difference of the two tagging jets is plotted for the \zzjj and
\zajj production processes. Note the similar phase space dependence. 
Here, the large K-factors in the tails might indicate that the
central scale at LO is too high. 
NLO curves for our three different scale choices (not shown) are
within the scale uncertainty band of a variation by a factor 2, 
highlighting the relevance of the NLO predictions.

Next, we investigate the radiative photon emission off the charged leptons in
\zajj and \wajj production. This radiative decay represents a simple QED
effect, which diminishes the sensitivity to anomalous couplings.  
Radiative decays 
dominate in the phase-space region where the reconstructed invariant mass of
the $Z\gamma$ ($W\gamma$) system is close to the $Z$ ($W$) mass. Thus, imposing
a cut on $M_{Z\gamma}$ $(M_{T,W\gamma})$ slightly above the $Z$ ($W$) mass should
remove these contributions.  %
For \zajj production, we can make use of the $pp\to \nu \bar{\nu} \gamma jj$
(``$Z_{\nu} \gamma jj$ '') channel, in which radiative decays are absent, to
determine the optimum value of the cut. The integrated cross section for the
``$Z_{l} \gamma jj$'' and ``$Z_{\nu} \gamma jj$'' channels are plotted as
functions of the $m_{Z\gamma}$ cut in the left panel of Fig.~\ref{fig:finalrad}. 
The latter is normalized by the ratio of the charge-lepton over the neutrino
branching ratio of the $Z$ boson. In the bottom panels, the ratio of the 
renormalized neutrino cross section to the electron cross sections are
plotted. 
Note the sharp
decrease of the cross section around the Z peak, indicating that radiative 
decays are eliminated. In the middle panel, one
observes that a cut around $m_{Z\gamma}^{\text{cut}}=120 \GeV$ would be
optimal -- the K-factor of the charge-lepton case stabilizes and equals the
neutrino channel. This is corroborated in the right panel, where the
normalized differential distributions of the reconstructed rapidity-azimuthal
angle separation of the $Z\gamma$ system are plotted for the two
channels. One observes in the K-factor panel that, for the curve with
the $M_{Z\gamma}> 120 \GeV$ cut, both channels behave 
approximately equal up to values of around 3, where the
different cuts applied to the leptons appear to have an effect.

For the \wajj production process, this comparison is not possible. In
Ref.~\cite{Campanario:2014dpa}, we showed that a cut on the transverse 
cluster mass
of the EW system of around $90 \GeV$ completely removes the radiative photon
emission off the charged leptons with a very mild reduction of the total cross
section of around $15\%$ (see Fig.~\ref{fig:finalradWA}).

\section{Conclusions}
\label{sect:con}
In these proceedings, NLO QCD results for several di-boson plus two jets
production processes at the LHC have been discussed. The NLO QCD corrections
significantly reduce the scale uncertainties. Large corrections can occur in
differential distributions. Different (reasonable) scale choices at NLO QCD
agree well with each other. However, large differences show up if only LO
 predictions are used.  These facts emphasize the necessity and the
relevance of the NLO QCD theoretical predictions. We have also found efficient
cuts which reduce the photon-radiated-off-lepton contributions for \wajj
and \zajj production.  All the processes are or will be available in the public program
{\texttt{VBFNLO}}.

\section{Acknowledgments}
FC acknowledges financial support by the IEF-Marie Curie program
(PIEF-GA-2011-298960) and partial funding by the LHCPhenonet
(PITN-GA-2010-264564) and by the MINECO (FPA2011-23596).  MK was funded by the
graduate program GRK 1694: ``Elementary particle physics at highest energy and
precision''. LDN and DZ are supported in part by the Deutsche
Forschungsgemeinschaft via the Sonderforschungsbereich/Transregio SFB/TR-9
``Computational Particle Physics''.




\nocite{*}
\bibliographystyle{elsarticle-num}
\biboptions{sort&compress}
\bibliography{QCDVVjj_p}






\end{document}